# Dynamic brain spectrum acquired by a real-time ultra-spectral imaging chip with reconfigurable metasurfaces

JIAN XIONG,[1,†] XUSHENG CAI,[1,†] KAIYU CUI,[1,2,†,*] YIDONG HUANG,[1,2] JIAWEI YANG,[1] HONGBO ZHU,[1] WENZHENG LI,[3] BO HONG,[3,4] SHIJIE RAO,[1] ZEKUN ZHENG,[1] SHENG XU,[1] YUHAN HE,[1] FANG LIU,[1,2] XUE FENG,[1,2] AND WEI ZHANG[1,2]

[1]*Department of Electronic Engineering, Tsinghua University, Beijing, 100084, China*

[2]*Beijing National Research Center for Information Science and Technology (BNRist), Tsinghua University, Beijing 100084, China*

[3]*Department of Biomedical Engineering, School of Medicine, Tsinghua University, Beijing, 100084, China;*

[4]*Tsinghua Laboratory of Brain and Intelligence (THBI), Tsinghua University, Beijing, 100084, China.*

**Corresponding author: kaiyucui@tsinghua.edu.cn*

**Abstract:** Spectral imaging paves way for various fields and particular in biomedical research. However, spectral imaging mainly depending on spatial or temporal scanning, cannot achieve high temporal, spatial and spectral resolution simultaneously. In this study, we demonstrated a silicon real-time ultra-spectral imaging chip based on reconfigurable metasurfaces, comprising of 155,216 (356×436) image-adaptive micro-spectrometers with ultra-high center-wavelength accuracy of 0.04 nm and spectral resolution of 0.8 nm. It is employed for imaging brain hemodynamics, and the dynamic spectral absorption properties of deoxyhemoglobin and oxyhemoglobin in a rat barrel cortex were obtained, which enlighten the spectroscopy in vivo studies and other real-time applications.

## 1. Introduction

Spectral imaging technology captures both spatial and spectral information for all points in the field of view [1,2], and has been applied in various fields, such as health, remote sensing, military, environmental monitoring [3], mining and geology [4], agriculture [5], and astronomy [6]. Real-time spectral imaging (RTSI) has shown great potential in biomedical research. The cortical spectral imaging technique has been used to study brain behavior and cortical activity patterns by continuously recording the hemodynamic responses to reveal the underlying mechanism of complex brain functions [7-13]. However, because only a few wavelength band data can be acquired continuously for the indicators [14], it is necessary to develop a broad spectral imaging method with high temporal and spatial resolutions for dynamic brain spectra. Nonetheless, spectral imaging still mainly depends on spatial or temporal scanning [15-20], which cannot achieve high temporal, spatial, and spectral resolution simultaneously.

In recent years, spectral devices based on micro/nano filters have been used to improve the integration and miniaturization of on-chip spectral devices. There are two main types of micro-nano filters: resonant filters and broadband filters. Resonant filters, such as micro-ring resonators [21-23], optical microcavities [24,25], and resonant metasurface structures [26-28], perform spectral analysis by filtering light of different wavelengths separately, and offer a relatively high spectral resolution. However, it is difficult to simultaneously produce a broad spectrum and high resolution using resonant filters considering the number of spectral channels corresponds to the number of filters. In contrast, broadband filters, such as quantum dot arrays [29], photonic crystal plate arrays [30,31], disordered scattering structures [32], and nanowires with tunable band gaps [33] encode the spectral information of incident light into the response of a set of filters at different detector positions, and uses a computational spectral algorithm to reconstruct the incident spectrum [29-35]. Because spectral information can be reconstructed for multiple wavelength points using fewer filter structures, it makes developing a micro spectrometer possible. However, despite extensive research on the potential of integrated micro-spectrometers to replace current spectrometers with complex structures and large volumes [29,30,32,33,36], the RTSI requirements could not be met. Therefore, developing new mechanisms to implement RTSI to analyze highly complex brain activities and various practical applications is still a big challenge.

In this study, we fabricated an RTSI chip based on a reconfigurable metasurface supercell on a CMOS image sensor (CIS) for imaging brain hemodynamics. Several locally distributed metasurface units can be dynamically combined and multiplexed in the metasurface supercell to reconfigure image-adaptive micro-spectrometers. As a result, we realized 155,216 ($356\times436$) micro-spectrometers on a CIS chip no larger than 0.5 $cm^2$ with an ultra-high center-wavelength accuracy of 0.04 nm and spectral resolution of 0.8 nm. The RTSI chip was used for imaging a rat's brain, which continuously monitored the dynamic absorption properties of deoxyhemoglobin and oxyhemoglobin over different wavelength bands, which are the indicators of neural activities. The result showed that the proposed chip is a promising method for studying brain functions pertaining to hemodynamics. Additionally, this scheme of reconfigurable metasurfaces allows the device to be directly extended to almost any commercial camera to seamlessly switch the information between the image and spectral image. Furthermore, spectral reconstruction can be easily combined with image recognition, which is significant in practical real-time applications.

## 2. Design of the proposed RTSI chip with reconfigurable metasurfaces

As shown in Figs. 1a and 1b, the proposed RTSI chip was designed by integrating a reconfigurable metasurface supercell on top of a CIS. All metasurface units (Fig. 1b) on the supercell can be dynamically combined and multiplexed to form thousands of image-adaptive micro-spectrometers (shown inset of Fig. 1a) using an image-adaptive processing algorithm (Supplementary S5), resulting in the best tradeoff between spatial resolution and spectral density. Additionally, the metasurface units operated as polarization-independent broadband filters to realize incident light modulation for all microspectrometers simultaneously, thereby ensuring the real-time imaging of the CIS, as shown in Figs. 1c and 1d. Moreover, the spectrum of incident light can be modulated with a high degree of freedom using the proposed reconfigurable metasurface supercell (Fig. 1d and Supplementary S1). Lastly, the output signals from the different modulation regions are collected by the CIS chip, and the incident spectrum is reconstructed by grafting the compressed sensing algorithm.

As many as 155,216 ($356\times436$) micro-spectrometers can be formed for the demonstrated RTSI chip using the reconfigurable supercell, where 400 metasurface units with different patterns form one base cell and 396 such base cells are arrayed as $18\times22$, as shown in Fig. 2. As a result, 155,216 micro-spectrometers can obtain spectral information at one time using the CIS, demonstrating the high temporal and spatial resolution of the device.

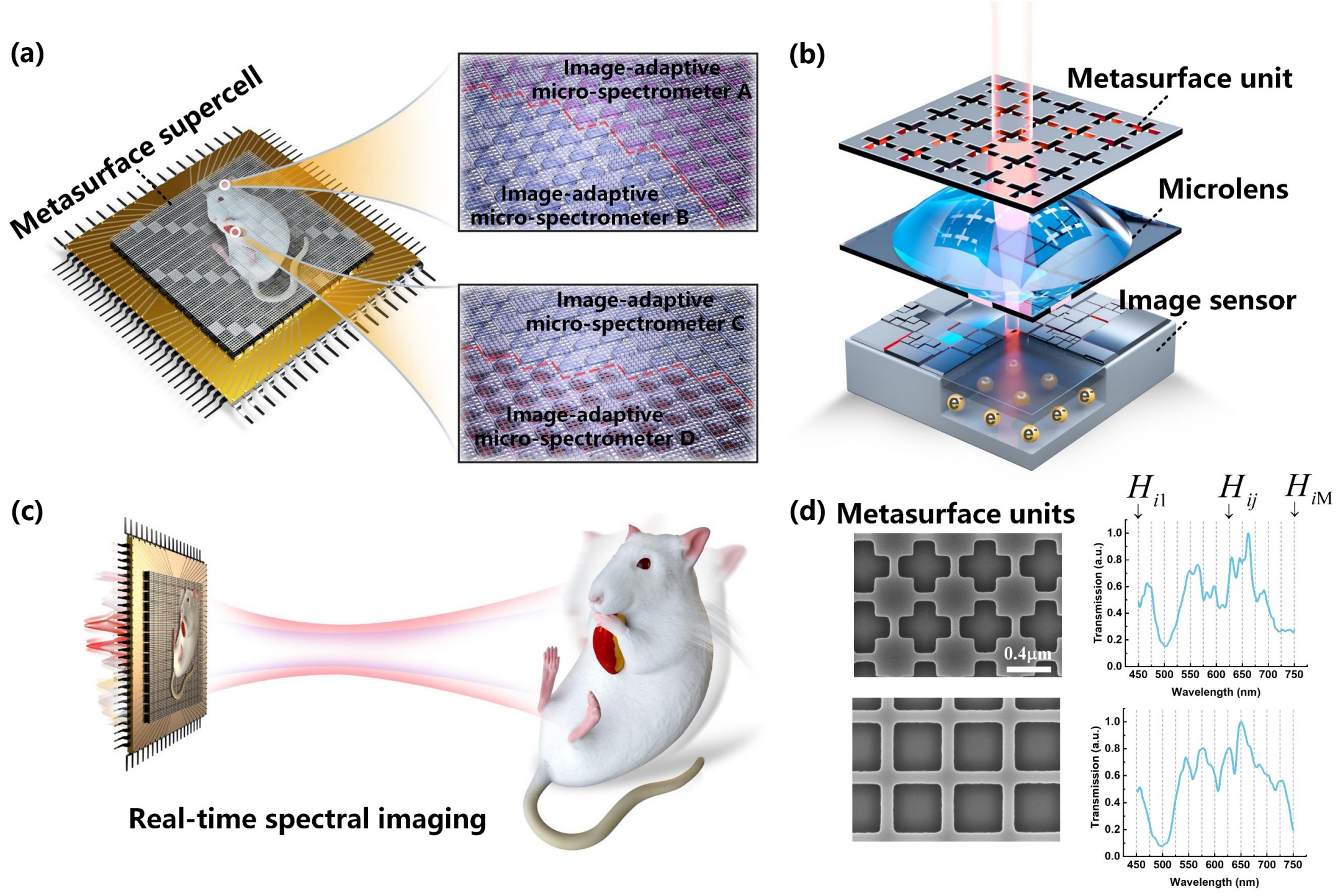

**Fig. 1.** Images showing the operation of the proposed ultraspectral imaging device. **a.** The device comprises a reconfigurable metasurface supercell with 158,400 metasurface units, and can form up to 155,216 micro-spectrometers when integrated with a commercial CMOS image sensor (CIS) chip (Thorlab DCC3260M). During the spectral reconstruction of an object, the adjacent metasurface units in the metasurface supercell are dynamically combined into a reconfigurable and image-adaptive micro-spectrometer, as shown in the zoomed-in figures, where the red and white areas represent two irregular shaped micro-spectrometer separated by the edges of the image. **b.** Schematic of the fundamental modulation unit, comprising the metasurface, microlens (used to improve quantum efficiency), and CMOS image sensor from top to bottom. **c.** Snapshot of spectral imaging. The light of the object to be imaged is incident on the metasurface supercell. **d.** SEM images of the selected metasurface units with $C_4$ symmetry. The corresponding modified transmission spectral curve is characterized using a monochromator for calibration. $H_{ij}$ denotes the modified transmission of the *i*-th metasurface at the *j*-th wavelength sampling point.

Electron beam lithography was used to form design patterns for the reconfigurable metasurface supercell (see details in Supplementary S3) on a 220 nm thick silicon-on-insulator (SOI) chip. The patterns were then transferred onto a silicon layer using inductively coupled plasma etching. Subsequently, the middle silicon dioxide layer was removed by wet etching using buffered hydrofluoric acid, resulting in the suspension of the metasurface supercell. Lastly, the top silicon layer of the SOI comprising the suspended metasurface supercell was transferred on top of a CIS chip using the polydimethylsiloxane (PDMS) transfer approach (see details in Supplementary S3). The proposed device can be fabricated entirely using the CMOS-compatible processing technology, which reduces the production cost.

The intensities of modulated incident light were recorded by the underlying CIS chip in the reconfigurable metasurface supercell, as shown in Fig. 1c. Then, the recorded data was processed and reconstructed into the incident spectrum using an algorithm. Herein, we introduced the working principle of a single microspectrometer. The transmission response of the *i*-th metasurface unit is denoted as $h_i(\lambda)$, where $i = 1, 2, 3, ..., N$, $N$ is the number of different patterns in a base cell (Fig. 2a), $\lambda$ is the wavelength, and $f(\lambda)$ is the incident spectrum to be measured. The signal intensity $I_i$ received by the CIS below the $i$-th metasurface unit is expressed as:

$$I_i = \int_{\lambda_1}^{\lambda_2} f(\lambda) h_i(\lambda) R(\lambda) P(\lambda) d\lambda = \int_{\lambda_1}^{\lambda_2} f(\lambda) H_i(\lambda)\, d\lambda \tag{1}$$

where $I_i$ is the average signal intensity received by $3 \times 3$ CIS pixels below one metasurface unit to improve the signal-to-noise ratio (SNR) and the relaxed alignment requirements during fabrication, $R(\lambda)$ is the absorption quantum efficiency of the CIS for wavelength $\lambda$, $P(\lambda)$ is the dispersion curve of the lens imaging system, and $\lambda_1$ and $\lambda_2$ are the lower and upper limits of the incident spectral distribution, respectively.

We set $H_i(\lambda) = h_i(\lambda) R(\lambda) P(\lambda)$, which is the modified transmission spectral curve that can be predetermined through measurement, as shown in Fig. 1d (see Supplementary S1 for details). Furthermore, the integral equations were discretized to produce the matrix equation (Eq. 2), as shown in Fig. 2d.

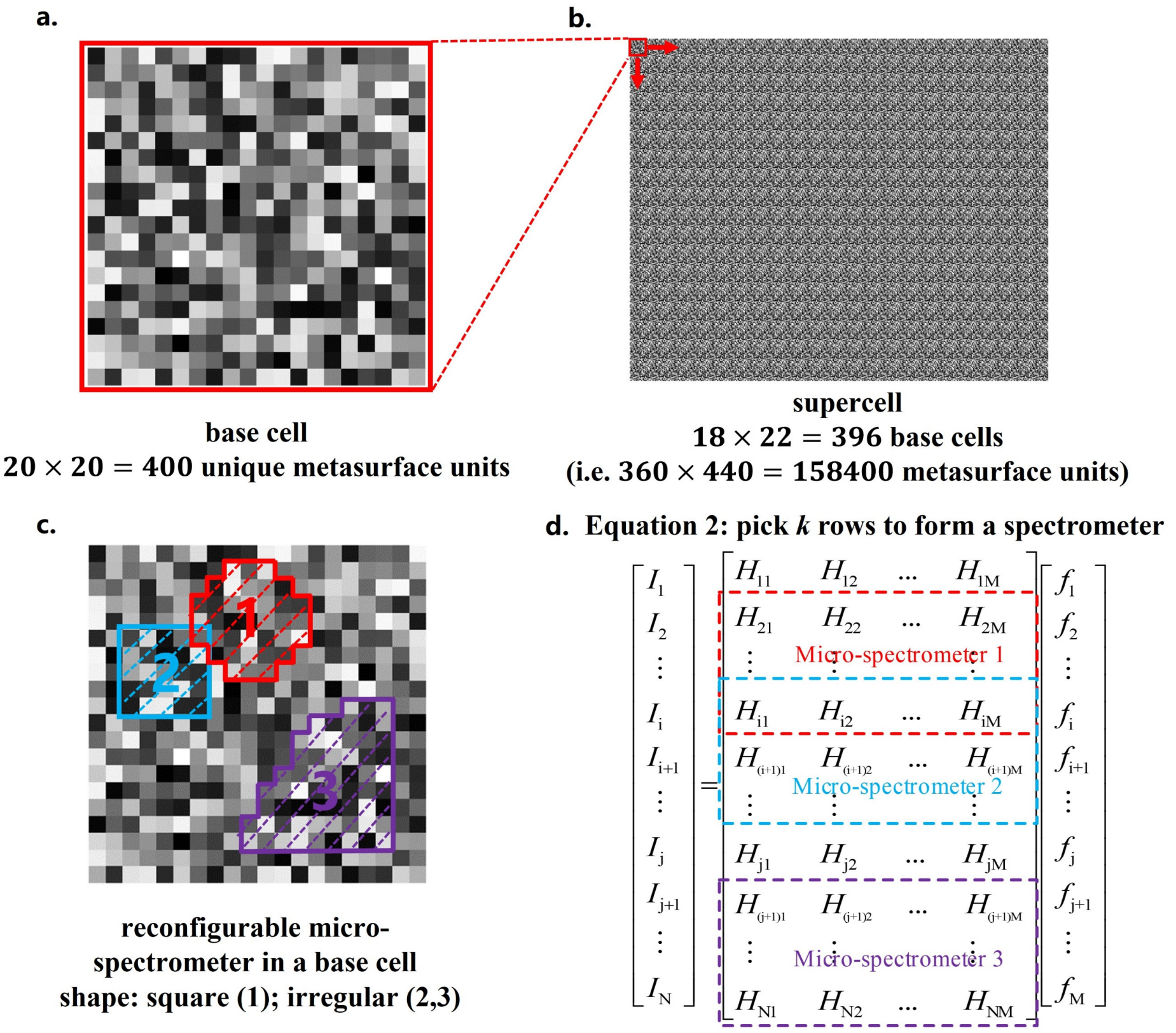

**Fig. 2.** Reconfiguration of the metasurface units. **a.** The base cell comprises $20 \times 20 = 400$ metasurface units. Pixels with different grayscale values represent the unique transmission spectrum of the metasurface units. **b.** The supercell comprises $18 \times 22 = 396$ base cells (i.e., $360 \times 440 = 158400$ metasurface units). The spectral imaging camera comprises a CMOS imaging sensor with the supercell chip attached to the surface. **c.** A reconfigurable micro-spectrometer can be constructed in any position on the base cell. The micro-spectrometer exhibits a flexible shape, and can be square (region 1) or irregular (region 2,3). Moreover, micro-spectrometers 1 and 2 share a common metasurface unit, thereby demonstrating the spatial multiplexing of metasurface units. In extreme cases of multiplexing, the entire supercell can be constructed into $356 \times 436$ micro-spectrometers. **d.** The matrix equation Eq. (2), discretized from the integral equation

Eq. (1). In the matrix, we select some rows to form the corresponding micro-spectrometers shown in Fig. 2c.

$I_i$ is the integral signal intensity of the *i*-th unit. Element $f_i$ in vector ***f*** denotes the value of the unknown spectrum ***f*** at the *i*-th wavelength band. Element $H_{ij}$ in matrix **H** denotes the transmission of the *i*-th metasurface at the *j*-th wavelength band (M equals the number of wavelength bands). Therefore, each row of the matrix **H** represents the transmission spectrum of a unique metasurface unit, whereas **H** represents a microspectrometer formed by the metasurface units. According to the theory of compressed sensing, **H** can be treated as a compressed sensing matrix [37] (i.e., CS matrix), used to sense the optical spectrum. In the proposed design, the transmission of every metasurface unit was simulated using a full-wave simulation software [38,39] to optimize the requirements of $C_4$ symmetry and compressed sensing [37] (see details in Supplementary S1).

Figs. 2c and 2d show the principle of a reconfigurable metasurface supercell. We assume that there exists N (N=400 in this study) metasurface units (corresponding to a base cell of the metasurface supercell) with varying transmissions in the region of interest. Considering the metasurface units are independent, we can select *k* number of units to form a CS matrix for spectral sensing. In this study, we used an image recognition technique to automatically determine the optimal shape of the microspectrometer, which can effectively avoid individual micro-spectrometer crossing image edges, where abrupt changes occur in the spectrum (see Supplementary S5). In addition, the multiplexing of metasurface can also be applied to improve the spatial resolution, as shown in Fig. 2c.

Finally, the algorithm of compressed sensing for the optical spectrum is grafted after setting up the optimal micro-spectrometers in the supercell. Dictionary learning based on sparse coding [40-42] was used to recreate the original spectrum (see Supplementary S4).

## 3. Experimental results

Fig. 3 shows the experimental results of the measurements of monochromatic light sources over a spectral range of 450-750 nm (at 0.5 nm intervals) with unit number *k*=25. According to the theory of compressed sensing, narrow-band spectra exhibit excellent reconstruction quality owing to their natural sparsity. Fig. 3a compares the reconstructed monochromatic lights from the single micro-spectrometer of the proposed device (blue line) and a commercial spectrometer (black dashed line, OceanView QE Pro). Fig. 3b shows that the center-wavelength accuracy of the monochromatic light was approximately 0.04 nm.

Considering the ultraspectral imaging chip can resolve two monochromatic lights with very similar wavelengths, we reduced the sampling interval to 0.2 nm and the spectral band to 2 nm (544.6-546.6 nm) to resolve ultra-fine spectral lines. A pair of peaks were formed using the 546 nm line of a mercury lamp and a tunable monochromatic light source. The micro-spectrometer effectively resolved the aforementioned double peaks with a wavelength interval of only 0.8 nm, as shown in Fig. 3c. Furthermore, the linewidths of the reconstructed double peaks (blue curves) were broadened to 0.23 and 0.24 nm. To the best of our knowledge, this double-peak resolution is the best result obtained for on-chip spectral imaging devices, and is approximately an order of magnitude higher than that obtained using a nanowire spectrometer (15 nm) [33]. Additionally, the acquired results of double-peak resolution surpass the results obtained using commercial portable spectrometers (OceanView QE Pro) by approximately 1.2 nm.

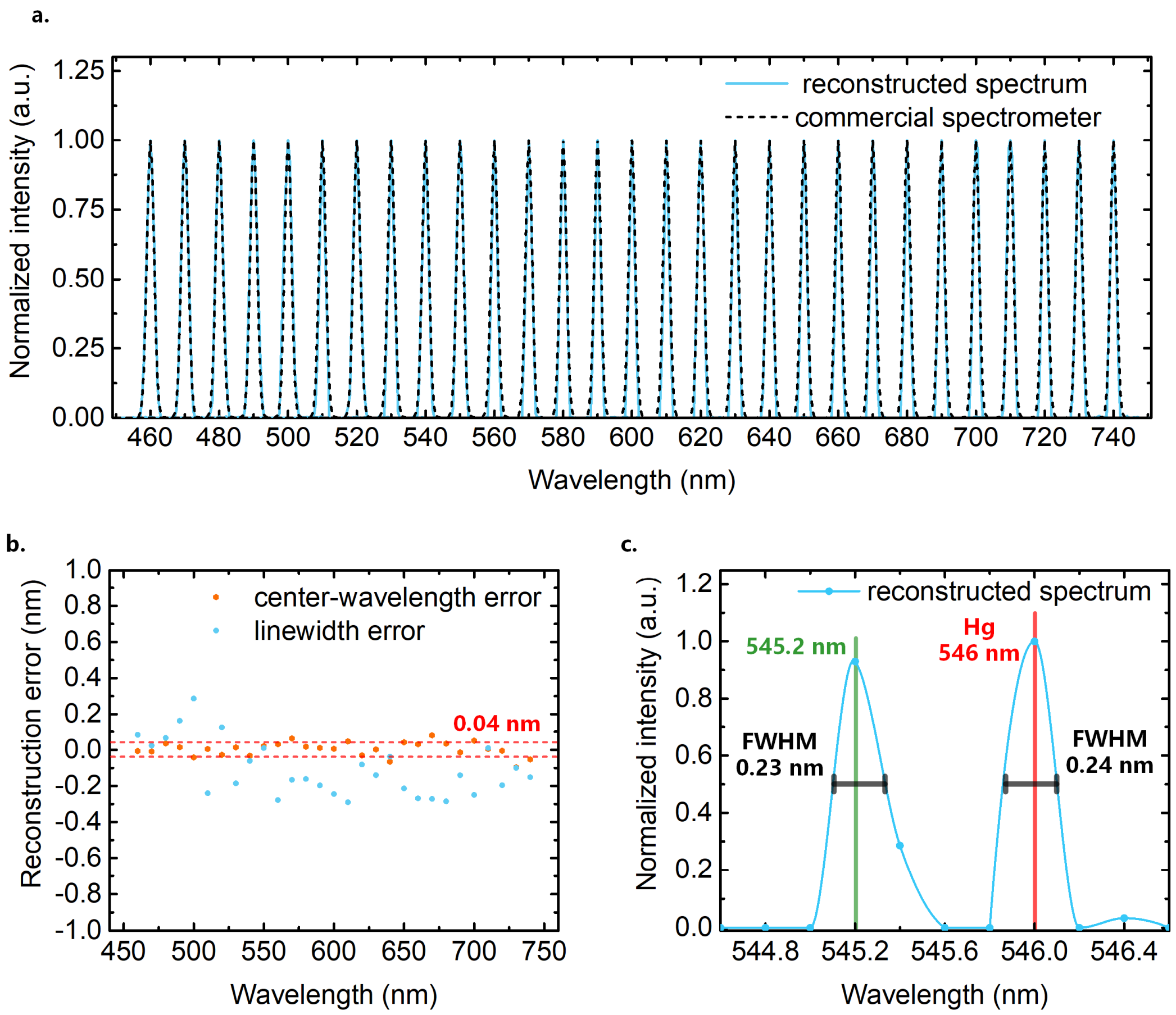

**Fig. 3.** Spectral reconstruction of the monochromatic light sources with $k$=25. **a.** Reconstruction results (450 - 750 nm, 0.5 nm intervals) of the narrow-spectrum lights obtained from the micro-spectrometer of the proposed device (blue line) and a commercial spectrometer, OceanView QE Pro (black dashed line). **b.** Errors in the reconstruction of the narrow-spectrum light. Here, we assumed the Gaussian envelope for the narrow-spectrum light to estimate the linewidth and center-wavelength. **c.** Reconstruction results of the double peaks corresponding to the 546 nm and 545.2 nm spectral lines of a mercury lamp and tunable monochromatic light source, respectively, with an interval of 0.8 nm between the two spectral lines. Other spectral lines of the mercury lamp are removed using filters during measurements.

Additionally, the proposed ultraspectral imaging device was able to accurately reconstruct complex broad-spectrum signals in the visible light region. Fig. 4 shows the reconstruction results for several types of broad spectra. The spectral range was maintained to 450 - 750 nm (0.5 nm interval). The blue and red curves were obtained using the micro-spectrometer of the proposed device and a commercial spectrometer, respectively, and served as a reference spectrum. The concept of fidelity was introduced to quantitatively compare the original and reconstructed spectra as:

$$F(X,Y) = \left(\sum_{\mathrm{m}} \sqrt{p_{\mathrm{m}} q_{\mathrm{m}}}\right)^2 \tag{3}$$

where $X$ and $Y$ are the normalized original and reconstructed spectra, respectively, and $p_{\mathrm{m}}$ and $q_{\mathrm{m}}$ are the corresponding intensities at the $m$-th wavelength sampling point. The fidelities of the broad-spectrum reconstruction obtained in this experiment were above 98%, even with a small number of metasurface units $k$=25, as shown in Figs. 4a–c, which indicates the good reconstruction ability of the proposed device.

Furthermore, we investigated the effect of the number of metasurface units *k* in a single micro-spectrometer on the fidelity of the reconstructed spectra to evaluate the spatial pixel reconstruction ability of the proposed device. *k* units were randomly selected from N=400 different metasurface patterns in the region of interest. The error bars represented the variance of the reconstruction fidelity obtained for multiple possible random combinations for each *k* value, as shown in Fig. 4d. As *k* increased, the reconstruction error decreased and the fidelity value increased, which was consistent with the principle of compressed sensing [37]. Moreover, the *k* value of a single micro-spectrometer can be dynamically adjusted depending on the reconfigurable metasurface supercell by considering the noise level for achieving an optimal trade-off between the pixel density and reconstruction quality of the reconstructed spectra. A small *k* value (large micro-spectrometer density), and a large *k* value (small micro-spectrometer density), can be used at low and high noise levels, respectively. Moreover, a high fidelity can still be realized with a *k* value as low as 25 for simple spectra, as shown in Fig. 4b.

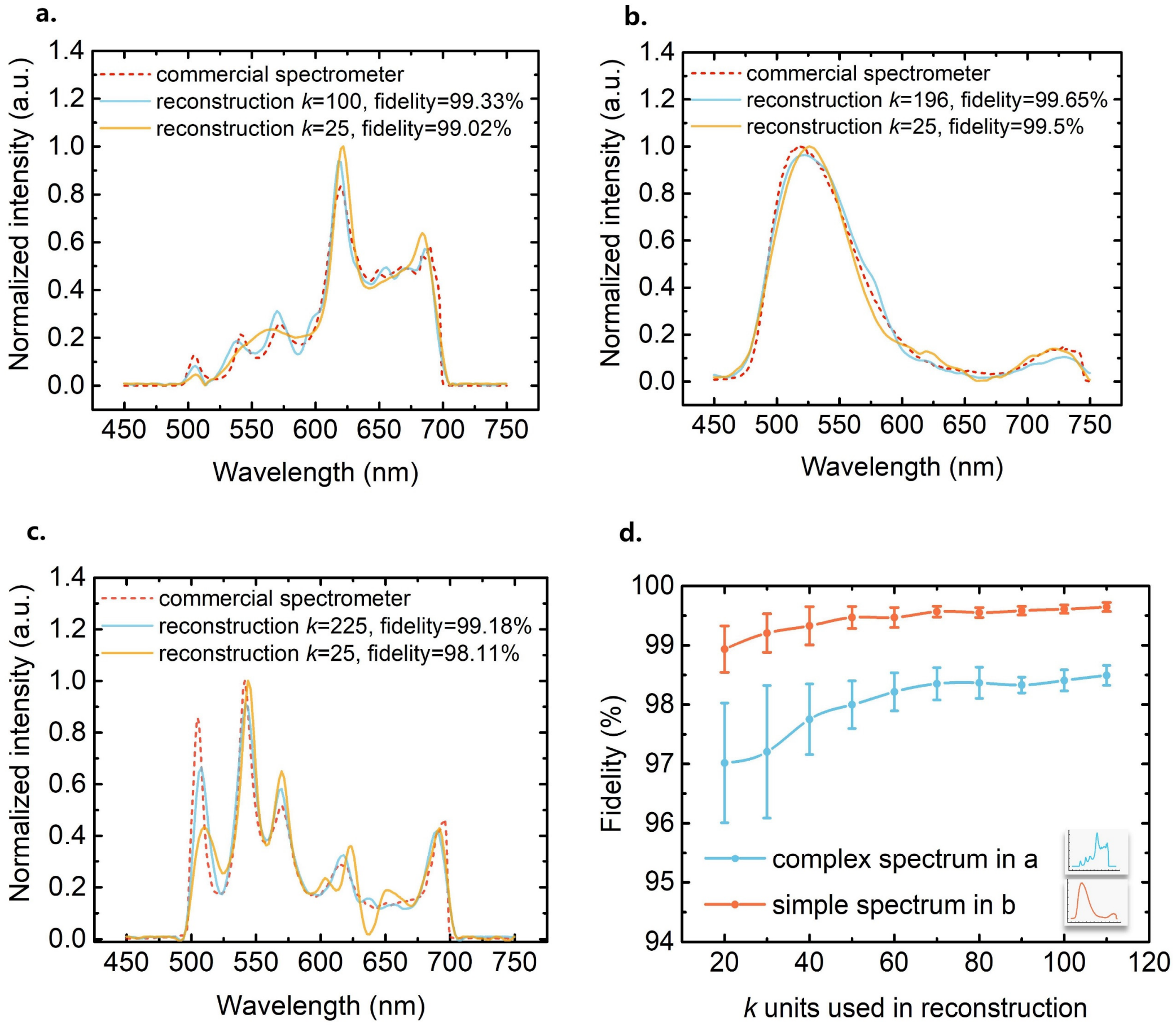

**Fig. 4. a, b, c.** Reconstruction results (450-750 nm, 0.5 nm interval) of the three broad spectra obtained using the micro-spectrometer of the proposed device (blue curve). The spectra detected by a commercial spectrometer are shown as reference (red dot-curve). Fidelities for the three different spectra are above 98%, even with a small number of metasurface units *k*=25, which indicates high consistency between the reconstructed and reference spectra. **d.** Spectrum-reconstruction fidelity of a micro-spectrometer with different unit number *k*. The error bars represent the variance in the fidelity obtained by randomly selecting *k* units. Herein, the fidelities of the reconstructed spectrum, as shown in Figs. 4a and b, are calculated as examples for a complex and simple spectra, respectively. For *k*=25, the fidelities deteriorated relatively, but showed good results with values larger than 96%. As *k* increases, the average fidelity gradually increases and variance decreases, which agrees with the theoretical analysis.

Lastly, we demonstrated an RTSI video of an *in vivo* rat brain using the fabricated device. Fig. 5 shows the RTSI results for the barrel cortex of the rat brain. The RTSI video (see Supplementary Video) was recorded using a microscope with a tenfold objective amplification, as shown in Fig. 5a (see Supplementary S6 for details). Herein, the original images, modulated images and multi-frame spectral images are shown respectively, where the pattern of micro-spectrometers is loomed in the latter and the cerebrovascular regions are denoted by red dashed lines. We obtained the ultraspectral information over a spectral range of 450 - 750 nm at a sampling interval of 0.5 nm using a snapshot for all points in the field of view. The spatial resolution was 87.9 μm and we set $k=100$ in the spectrum reconstruction to ensure spectral resolution. Moreover, we converted the spectrum data to a post-colored spectral image and a data cube, wherein we selected spectral images at five single wavelengths of 510, 540, 570, 600, and 630 nm. Fig. 5c shows the cerebrovascular spectral reconstruction results of the barrel cortex at selected regions, as shown in Fig. 5a, where regions 1, 2, and 5 represent the vascular regions and regions 3, 4, and 6 represent the non-vascular regions. Evidently, the absorption valleys of oxyhemoglobin (HbO, 577 nm), carboxyhemoglobin (HbCO, 569 nm), deoxyhemoglobin (HbR, 555 nm), and HbO/HbCO (540 nm) were significantly visible from the characteristic absorption data of hemoglobin [43,44] (see Supplementary S7 for details), thereby indicating the potential of the proposed RTSI chip for analyzing the hemoglobin concentration distribution *in vivo*, which is closely associated with brain function [8].

Figs. 5d and e show the curve of the characteristic spectral intensity change versus time in regions 5 and 6, respectively, which demonstrates the dynamic variation of hemoglobin. In the vascular area (region 5, Fig. 5d), the dominant process is the transport of hemoglobin, which results in a parallel change in the concentrations of hemoglobin derivatives (a positive correlation with a slope greater than 0). Meanwhile, in the non-vascular area (region 6, Fig. 5e), the dominant process is the exchange of oxygen between the capillaries and tissue cells, wherein the cells absorb oxygen and convert HbO to HbR. Therefore, the concentrations of HbO and HbR showed a negative correlation with slopes less than 0, which is consistent with our experimental results. The demonstrated spectral video of the barrel cortex can be used to monitor the excited brain regions and study them based on a millisecond-level real-time response. Therefore, the proposed RTSI chip can non-invasively analyze brain activity and open a new avenue for neuroscience and brain science.

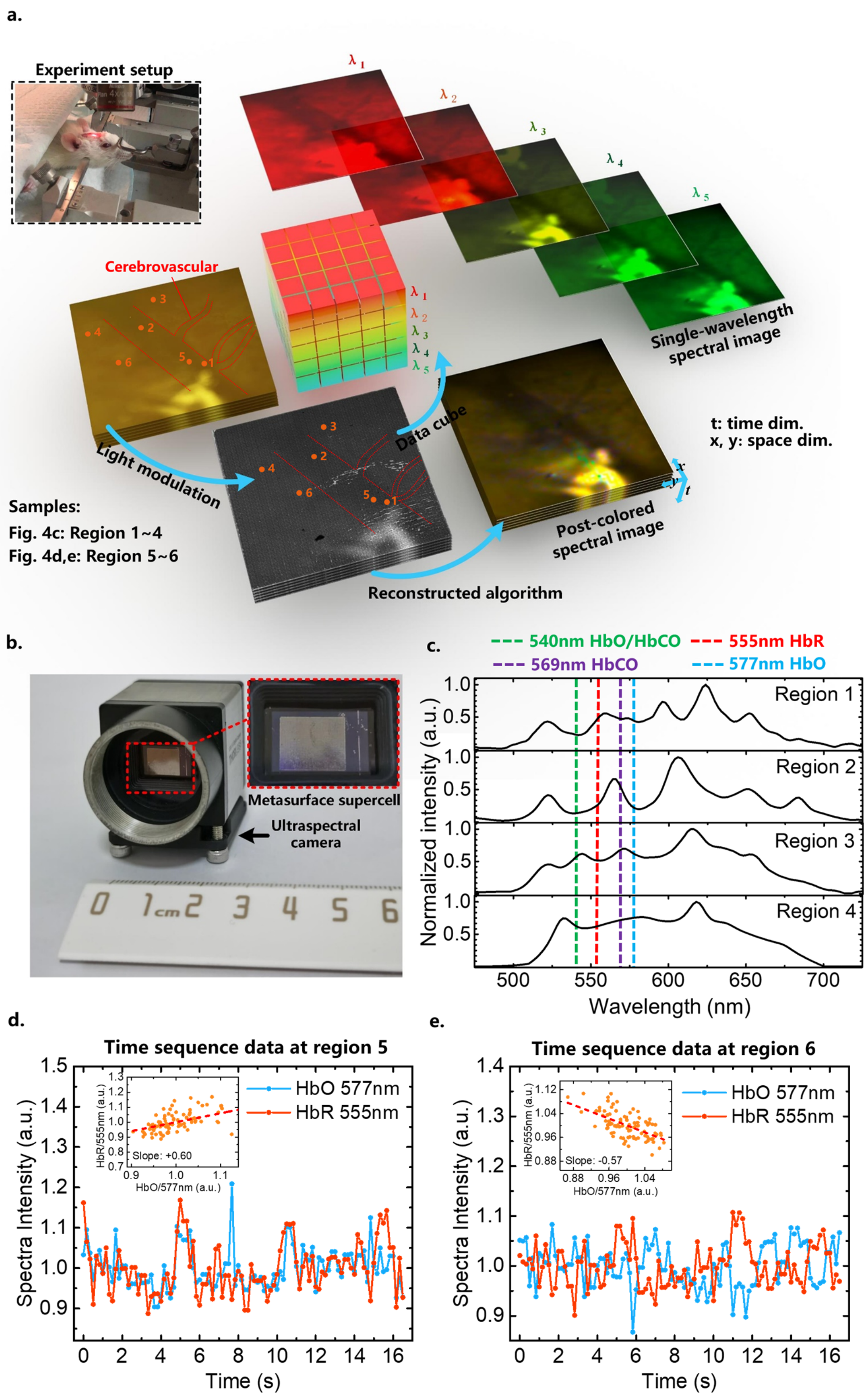

**Fig. 5. a.** Multi-frame images of rat cerebrovascular captured using an ordinary color camera and the proposed RTSI chip (black-and-white). A black-and-white photo was captured using an ordinary LED source. A post-colored spectral image and a data cube with 601 bands were computationally reconstructed from the spectral imaging data, wherein five single-wavelength spectral images were selected from the 601 bands. The chroma of the spectral image is reconstructed based on the Commission Internationale de l´Eclairage (CIE) 1931 color space [45]. The grayscale of the spectral image is calculated from the black-and-white photo to distinguish different regions of the rat brain. **b.** An ultraspectral imaging chip with a reconfigurable metasurface supercell is placed on top of the CIS and used in all experiments conducted in this study. **c.** Spectrum reconstruction at $k = 100$. The reconstructed absorption spectra of oxyhemoglobin (HbO, 577 nm), carboxyhemoglobin (HbCO, 569 nm), deoxyhemoglobin (HbR, 555 nm), and HbO/HbCO (540 nm) were significantly visible. **d** Measured spectral intensity data versus time in region 5 The two selected wavelengths (555 and 577 nm) correspond to the absorption peaks of HbO and HbR, respectively. The inset figure shows the positive correlation between HbO and HbR in the vascular area. **e.** Measured spectral intensity data versus time in region 6. The negative correlation between HbO and HbR in the non-vascular areas can be recognized.

## 4. Conclusion

This study proposed and demonstrated an ultraspectral imaging chip based on a reconfigurable metasurface supercell. Further, an image-adaptive strategy was employed to provide the best tradeoff between spatial and spectral resolutions. The micro-spectrometers of the proposed device exhibited a high center-wavelength accuracy of 0.04 nm, a high spectral resolution of 0.8 nm, and a broad wavelength range of 300 nm. Compared to the latest report of on-chip spectrometers [26,29,31,33], this spectral resolution for a single microspectrometer is considered the best, and is approximately an order of magnitude higher than that obtained using a nanowire spectrometer (15 nm resolution [33]), while the spatial resolution (87.9 μm) is maintained at the same magnitude (100 μm for nanowire spectrometer [33]). Meanwhile, the temporal resolution was improved from a scanning scenario (< 1 Hz [33]) to a real-time scenario (typical value 30 Hz for normal CIS, maximum value 1000 Hz for high-speed CIS). Therefore, the proposed reconfigurable chip can be a potential new method for RTSI with both high spectral and spatial resolutions.

Furthermore, we performed an *in vivo* experiment on a rat brain (barrel cortex region) to obtain real-time ultraspectral video with $\Delta\lambda/\lambda$~0.001 ($\lambda$: 450-750 nm, $\Delta\lambda$: 0.5 nm interval). The results show a correlation between the optical spectrum and hemoglobin concentration. Affected by some defects of the spectral chip and optical aberration of microscopic imaging systems, the quality of the video is not very high, which would be improved by further iterated experiments in order to eventually achieve the dynamic detection of brain hemoglobin concentration. Finally, we seamlessly integrated the reconfigurable metasurface supercell with commercial cameras to avoid system incompatibility issues and enable real-time dynamic spectrum measurement in all optical imaging systems.

**Data and materials availability:** The data supporting the findings of this study, including the data used for the plots and custom code, are available with the corresponding authors upon reasonable request.

**Acknowledgments:** The authors would like to thank Tianjin H-Chip Technology Group Corporation and Innovation Center of Advanced Optoelectronic Chip and Institute for Electronics and Information Technology in Tianjin, Tsinghua University, for their support during SEM and ICP etching. This study was supported by the Tsinghua University Initiative Scientific Research Program, Beijing Innovation Center for Future Chips, Tsinghua University, Beijing Frontier Science Center for Quantum Information, and the Beijing Academy of Quantum Information Sciences.

**Funding:** The National Key R&D Program of China (Contract No. 2018YFB2200402), National Natural Science Foundation of China (Grant Nos. 91750206 and 61775115), Beijing Municipal Science Technology Commission Z201100004020010, Beijing National Science Foundation contract Z180012,

Beijing Frontier Science Center for Quantum Information, and Beijing Academy of Quantum Information Sciences.

**Disclosures:** The authors declare no conflicts of interest.

**Author contributions:** J.X., X.C., and K.C. contributed equally to this study. X.C. proposed and evaluated the characteristics of micro-spectrometers. K.C. conceived the study and proposed a spectral-imaging strategy. J.X. designed the metasurface structures and conducted the spectral imaging experiments. B.H. and W.L. designed the experimental paradigm used in *the in vivo* experiments on rats. Y.H. supervised the project and advised the device optimization. J.X., K.C., Y.H., B.H., and W.L. wrote the manuscript with contributions from all other coauthors. H.Z. and Z. Z. participated in the fabrication process. S.X. and Y.H. participated in the algorithm optimization. F.L., X.F., and W.Z. provided useful comments on the results. All authors read and approved the manuscript.